\documentclass[reprint,amsmath,amssymb,prl,aps,]{revtex4-2}
\usepackage{graphicx}
\usepackage{dcolumn}
\usepackage{bm}
\usepackage{color}
\usepackage{braket}
\usepackage[dvipdfmx]{hyperref}
\begin{document}
\preprint{APS/123-QED}
%%%%%%%%%%%%%%%%%%%%%%%%%%%%%%%%%%%%%%%%%%%%%%%%%%%%%%%%%%%%%%%%
\title{Fingerprints of possible even-parity superconducting states in Sr$_2$RuO$_4$ \\ detected by planar tunneling spectroscopy}
\author{Satoshi Ikegaya$^{1,2}$, Shu-Ichiro Suzuki$^{2}$, Yukio Tanaka$^{2}$, and Dirk Manske$^{1}$}
\affiliation{$^{1}$Max-Planck-Institut f\"ur Festk\"orperforschung, Heisenbergstrasse 1, D-70569 Stuttgart, Germany\\
$^{2}$Department of Applied Physics, Nagoya University, Nagoya 464-8603, Japan}
\date{\today}
%%%%%%%%%%%%%%%%%%%%%%%%%%%%%%%%%%%%%%%%%%%%%%%%%%%%%%%%%%%%%%%%
\begin{abstract}
After more than 25 years of research, three even-parity superconducting states---the $d+id$-wave, $d+ig$-wave, and $s+id$-wave states---have emerged as leading candidates for the superconducting states of Sr$_2$RuO$_4$.
In the present work, we propose a tunneling spectroscopy experiment for distinguishing among these three superconducting states.
The key component of our proposal is that we examine the conductance spectra of normal-metal/Sr$_2$RuO$_4$ junctions with various angles between the junction interface and the crystal axis of the Sr$_2$RuO$_4$.
The angle dependence of the conductance spectra shows a unique pattern in each superconducting state, which can function as a fingerprint for verifying the pairing symmetry of Sr$_2$RuO$_4$.
\end{abstract}
%%%%%%%%%%%%%%%%%%%%%%%%%%%%%%%%%%%%%%%%%%%%%%%%%%%%%%%%%%%%%%%%%%%
\maketitle

\section{Introduction}
At present, the physics of the superconductor Sr$_2$RuO$_4$ (SRO)~\cite{maeno_03,kallin_09,maeno_17}, discovered in 1994~\cite{maeno_94}, faces an important turning point.
Specifically, recent precise experiments involving the nuclear magnetic resonance Knight shift~\cite{brown_19,ishida_20,brown_20} and polarized neutron scattering~\cite{hayden_20} appear to be inconsistent with a spin-triplet chiral superconducting state in this compound,
which has heretofore been considered to have been realized.
Motivated by these impactful observations, researchers have intensively investigated other possible superconducting states in SRO.
In particular, three even-parity superconducting states---$d_{zx}+id_{yz}$-wave~\cite{agterberg_20(r)}, $d_{x^2-y^2}+ig_{xy(x^2-y^2)}$-wave~\cite{thomale_20}, and $s+id_{xy}$-wave states~\cite{kee_21}---have attracted attention because they can explain the results of two recent experiments: (i) ultrasound experiments in which discontinuous jumps in the $c_{66}$ shear modulus at the superconducting transition temperature were observed ~\cite{proust_20,ramshaw_20} and (ii) muon spin relaxation experiments under uniaxial strains, where splits were observed between the superconducting transition temperature and the temperature of the onset of broken time-reversal symmetry~\cite{grinenko_20}.
The design of experiments to distinguish among these three possible superconducting states is currently an active research topic in this field.

\begin{figure}[hhhh]
\begin{center}
\includegraphics[width=0.45\textwidth]{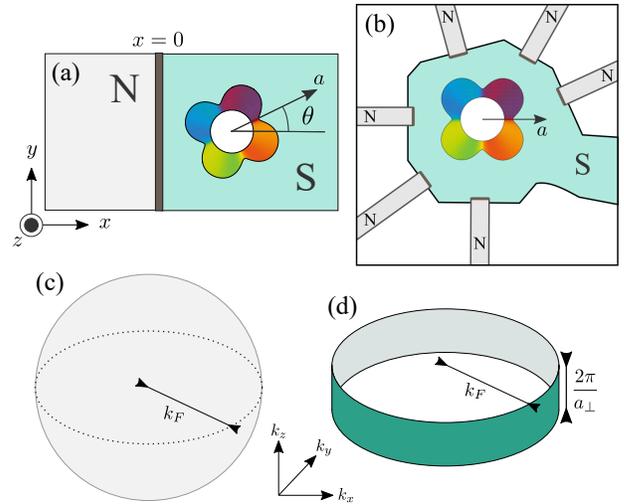}
\caption{(a) Schematic (top view) of the NS junction.
The angle between the $a$-axis of Sr$_2$RuO$_4$ and the normal vector of the junction interface is given by $\theta$.
(b) Possible experimental setup for examining the $\theta$-dependence of the differential conductance.
(c) Spherical Fermi surface for the normal-metal segment.
(d) Cylindrical Fermi surface for the superconducting segment.
}
\label{fig:figure1}
\end{center}
\end{figure}

In the present paper, we discuss the planar tunneling spectroscopy of normal-metal/superconductor (NS) junctions, where the superconducting segment is characterized by the $d_{zx}+id_{yz}$-wave, $d_{x^2-y^2}+ig_{xy(x^2-y^2)}$-wave, and $s+id_{xy}$-wave pairing symmetries.
A key component of our scheme is that we examine the conductance spectra acquired at various angles between the junction boundary and the crystal axis of SRO [see Figs.~\ref{fig:figure1}(a) and \ref{fig:figure1}(b)].
We demonstrate that each superconducting state has a characteristic pattern in the conductance spectra with respect to the junction angle, which can be used as a fingerprint for verifying the superconducting state of SRO (see Fig.~\ref{fig:figure3}).
Notably, a similar strategy has been used to demonstrate the $d$-wave nature of high-$T_c$ cuprate superconductors~\cite{kashiwaya_94,kajimura_97,wei_98,kashiwaya_00(2),kashiwaya_00}.
In addition, tunneling junctions of SRO have been fabricated in numerous experiments thus far~\cite{laube_00,upward_02,flouquet_09,kambra_11,kashiwaya_11,yada_14,firmo_15,mao_15,wei_17}.
In particular, the \textit{planar} tunneling spectroscopy of SRO with certain junction angles has already been demonstrated in experiments~\cite{kashiwaya_11,yada_14}.
The proposed experiments would therefore be implemented using current technologies.
We thus propose a useful and practical experiment for identifying the pairing symmetry of the superconductor SRO.
%------------------------------------------------------------------------------
%------------------------------------------------------------------------------

\section{Model and Formulation}%
We here consider a three-dimensional NS junction such as that shown in Fig.~\ref{fig:figure1}(a), where the relative angle between the normal vector of the junction interface and the $a$-axis of SRO is denoted by $\theta$.
The junction interface is located at $x=0$, and we apply a periodic boundary condition in the $y$ and $z$ directions.
Because a microscopic model of SRO has not yet been established, we instead employ a simple single-band model that focuses only on the $\gamma$-band of SRO.
The superconducting segment for $x>0$ is described by the Bogoliubov--de Gennes Hamiltonian,
\begin{align}
&H_S(\boldsymbol{k}) = \left[ \begin{array}{cc}
\xi_S(\boldsymbol{k}) & \Delta(\boldsymbol{k}) \\ \Delta^{\ast}(\boldsymbol{k}) & -\xi_S(\boldsymbol{k}) 
\end{array} \right], \\
&\xi_S(\boldsymbol{k}) = \frac{\hbar^2}{2m} (k_x^2 + k_y^2) - \mu_F.
\end{align}
We assume the absence of kinetic energy with respect to $k_z$ so that the SRO has a cylindrical Fermi surface, as shown in Fig.~\ref{fig:figure1}(c).
The radius of the cylinder is $k_F=\sqrt{2m\mu_F}/\hbar$.
The height of the cylinder is given by $2 \pi/a_{\perp}$, where $a_{\perp}$ represents the lattice constant along the $c$-axis of SRO.
We vary $k_z$ in the range $-\pi/a_{\perp} < k_z < \pi/a_{\perp}$.
The specific form of the pair potential $\Delta(\boldsymbol{k})$ is discussed later.
The normal segment for $x<0$ is described by
\begin{align}
&H_N(\boldsymbol{k}) = \left[ \begin{array}{cc}
\xi_N(\boldsymbol{k}) & 0 \\ 0 & -\xi_N(\boldsymbol{k})
\end{array} \right], \\
&\xi_N(\boldsymbol{k}) = \frac{\hbar^2}{2m} (k_x^2 + k_y^2+k_z^2) - \mu_F.
\end{align}
The normal segment has a spherical Fermi surface of the radius $k_F$, as shown in Fig.~\ref{fig:figure1}(b).
In addition, we consider a potential barrier $V_0 \delta(x)$ at the junction interface.
We evaluate the differential conductance of the NS junction using the formula~\cite{klapwijk_82,bruder_90,tanaka_95,tanaka_96,sengupta_02}
\begin{align}
G = \frac{2e^2}{h} \int_{-\alpha}^{\alpha}dk_z \int_{-\beta}^{\beta}dk_y
 \left[ 1- |r^{ee}|^2 + |r^{he}|^2\right]_{E=eV},
\end{align}
where $\alpha=\pi/a_{\perp}$ and $\beta=\sqrt{k_F^2-k_z^2}$.
The normal and Andreev reflection coefficients at energy $E$ are denoted as $r^{ee}$ and $r^{he}$, respectively.
These reflection coefficients are determined by the boundary conditions:
\begin{align}
&\varphi_N(x=0) = \varphi_S(x=0),\\
&\partial_x \varphi_N(x)|_{x=0} = \partial_x \varphi_S(x)|_{x=0} - \frac{2m V_0}{\hbar^2} \varphi_S(x=0),
\end{align}
where $\varphi_{N(S)}$ represents the wave functions in the normal (superconducting) segment.
By assuming $E, \Delta_0 \ll \mu_F$, we obtain
\begin{align}
&\varphi_N(x) = \left[ \begin{array}{c} 1 \\ r^{he} \end{array} \right] e^{ik_Nx}
+ \left[ \begin{array}{c} r^{ee} \\ 0 \end{array} \right] e^{-ik_Nx},\\
&k_N = \sqrt{k_F^2-k_y^2-k_z^2},
\end{align}
and
\begin{align}
&\varphi_S(x) = t^{ee}\left[ \begin{array}{c} u_+ \\ v^{\ast}_+ \end{array} \right] e^{ik_Sx}
+ t^{he}\left[ \begin{array}{c} v_- \\ u_- \end{array} \right] e^{-ik_Sx},\\
& u_{\pm} = E + \Omega_{\pm}, \quad \Omega = \sqrt{E^2-\Delta_{\pm}},\\
& v_{\pm} = \Delta_{\pm} = \Delta(\pm k_S,k_y,k_z),\\
&k_S = \sqrt{k_F^2-k_y^2},
\end{align}
where we ignore the insignificant terms $O\left( \frac{E}{\mu_F},\frac{\Delta_0}{\mu_F}\right)$.

In the present paper, we study the differential conductance of the junction in Fig.~\ref{fig:figure1}(a) with various junction angles $\theta$.
In experiments, the $\theta$-dependence of the differential conductance can be examined using, for instance, a device with multiple junctions fabricated on the same SRO chip, as shown in Fig.~\ref{fig:figure1}(b).
Similar techniques have been used in tunneling spectroscopy experiments involving high-$T_c$ cuprate superconductors~\cite{kashiwaya_00(2)}.

%------------------------------------------------------------------------------
\begin{figure}[tttt]
\begin{center}
\includegraphics[width=0.4\textwidth]{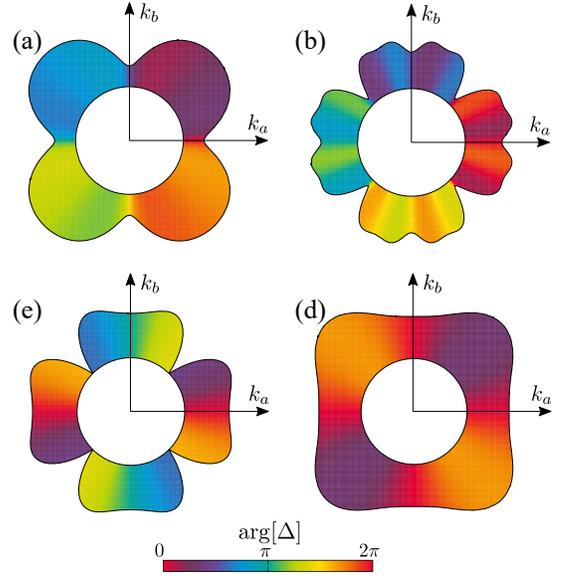}
\caption{Pair potentials of the possible superconducting states in Sr$_2$RuO$_4$:
(a) $d_{zx}+id_{yz}$-wave, (b) $(d_{zx}+id_{yz})^{\prime}$-wave,
(c) $d_{x^2-y^2}+ig_{xy(x^2-y^2)}$-wave, and (d) $s+id_{xy}$-wave pair potentials.
The boundary between the white circle and the colored regions represents the Fermi surface.
The distance between the Fermi surface and the outer boundary of the colored region is proportional to the amplitude of the pair potential at the Fermi surface, i.e., $|\Delta(k)|$.
The color corresponds to the phase of the pair potential, i.e., $\mathrm{arg}[\Delta(k)]$.
In (a) and (b), we chose $k_z = 0.5\pi$. The pair potentials in (c) and (d) are independent of $k_z$.}
\label{fig:figure2}
\end{center}
\end{figure}
%------------------------------------------------------------------------------
We now consider the specific form of the pair potentials.
To describe the $d_{zx}+id_{yz}$-wave state, we employ two different pair potentials.
The first pair potential contains only the shortest-range pairing belonging to $d_{zx}+id_{yz}$-wave pairing symmetry:
\begin{align}
\Delta(\boldsymbol{k}) = \Delta_0 \left( \sin k_a a_{\parallel} + i \sin k_b a_{\parallel} \right) \sin k_z a_{\perp},
\label{eq:pair1}
\end{align}
where
\begin{align}
&k_a = k_F \left( \hat{k}_x \cos \theta + \hat{k}_y \sin \theta \right),\\
&k_b = k_F \left( -\hat{k}_x \sin \theta + \hat{k}_y \cos \theta \right),
\end{align}
with $\hat{k}_{x(y)}=k_{x(y)}/\sqrt{k_x^2+k_y^2}$.
Here, $k_{a(b)}$ represents the momenta along the $a$-axis ($b$-axis) of SRO and $\sqrt{k_a^2+k_b^2}=k_F$ is satisfied.
Parameter $a_{\parallel}$ denotes the lattice constant along the $a$-axis and $b$-axis of SRO.
As shown in Fig.~\ref{fig:figure2}(a), the pair potential in Eq.~(\ref{eq:pair1}) has the largest gap amplitudes along $k_a = \pm k_b$.
This observation is consistent with previously reported results of a specific heat experiment~\cite{deguchi_04}.
However, the results of a recent experiment involving quasiparticle interference patterns suggest the presence of gap nodes or gap minima along $k_a = \pm k_b$~\cite{madhavan_20}.
In addition, a recently proposed microscopic theory also predicts a $d_{zx}+id_{yz}$-wave state with gap minima along $k_a = \pm k_b$~\cite{agterberg_20(r)}.
Thus, to reproduce such gap minima within an effective single-band model, we phenomenologically consider a pair potential containing longer-range pairing components:
\begin{align}
\Delta(\boldsymbol{k}) =\Delta_0 \{ p_a(\boldsymbol{k}) + i p_b(\boldsymbol{k}) \} \sin k_z a_{\perp}, \label{eq:pair2}
\end{align}
with
\begin{align}
p_{a(b)}(\boldsymbol{k}) =& c_1 \sin k_{a(b)} a_{\parallel} \nonumber\\
&+ c_2 \cos  k_{b(a)} a_{\parallel} \times \sin k_{a(b)} a_{\parallel} \nonumber\\
&+ c_3 \sin 3k_{a(b)} a_{\parallel}.
\end{align}
Notably, similar long-range pairing terms with large amplitudes have been proposed in previous microscopic studies investigating chiral $p$-wave states of SRO~\cite{wang_13,simon_15}.
For simplicity, in the present work, we chose $c_1 = c_2 = c_3 = c$ and determined the coefficient $c$ under a condition of
$\mathrm{max}[p_{a}^2(\boldsymbol{k}) + p_{b}^2(\boldsymbol{k})]=1$.
As shown in Fig.~\ref{fig:figure2}(b), the resultant pair potential indeed exhibits the gap minima along $k_a = \pm k_b$.
We hereafter refer to the superconducting state described by the pair potential in Eq.~(\ref{eq:pair2}) as the $(d_{zx}+id_{yz})^{\prime}$-wave state.
%
%------------------------------------------------------------------------------
\begin{figure}[bbbb]
\begin{center}
\includegraphics[width=0.5\textwidth]{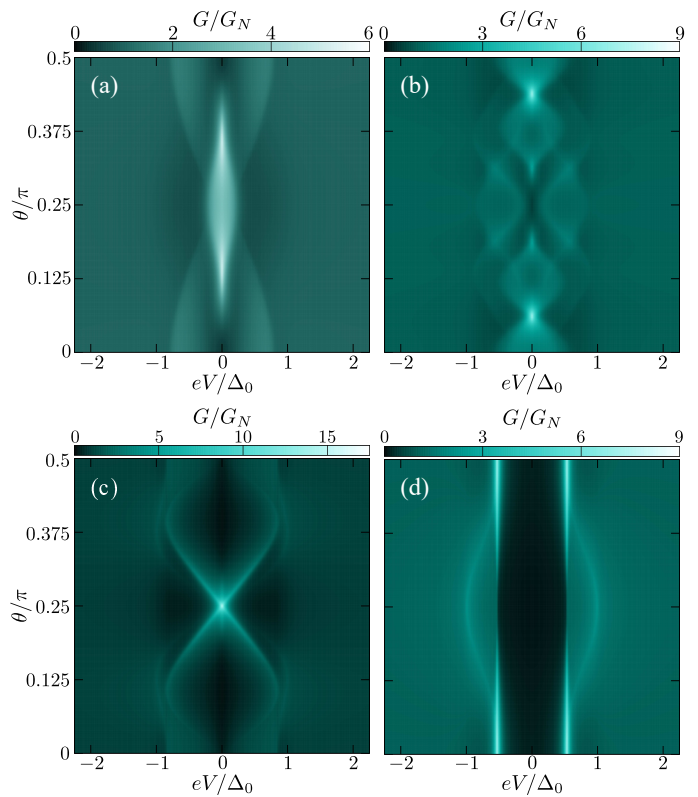}
\caption{Differential conductance as a function of the bias voltage and the relative angle between the $a$-axis of SRO and the normal vector of the junction interface $\theta$.
The differential conductance of the NS junction (i.e., $G$) is normalized by the normal conductance $G_N$.
We varied the junction angle $\theta$ in the range $0 \leq \theta \leq 0.5\pi$.
In (a)--(d), we consider (a) $d_{zx}+id_{yz}$-wave, (b) $(d_{zx}+id_{yz})^{\prime}$-wave, (c) $d_{x^2-y^2}+ig_{xy(x^2-y^2)}$-wave, and (d) $s+id_{xy}$-wave pair potentials.
}
\label{fig:figure3}
\end{center}
\end{figure}
%------------------------------------------------------------------------------
%
We then describe the $d_{x^2-y^2}+ig_{xy(x^2-y^2)}$-wave state by the pair potential,
\begin{align}
&\Delta(\boldsymbol{k}) = \Delta_0 \{ p_{d1}(\boldsymbol{k}) + i p_{g}(\boldsymbol{k})\},\label{eq:pair3}\\
&p_{d1}(\boldsymbol{k}) = c_{d1} ( \cos k_b a_{\parallel} - \cos k_a a_{\parallel} ),\\
&p_{g}(\boldsymbol{k}) = c_{g} \sin k_a a_{\parallel}\sin k_b a_{\parallel} ( \cos k_b a_{\parallel} - \cos k_a a_{\parallel} ).
\end{align}
The coefficients $c_{d1}$ and $c_{g}$ are determined by the conditions
\begin{align}
&\braket{p_{d1}^2(\boldsymbol{k})}_{\mathrm{FS}} = \braket{p_g^2(\boldsymbol{k})}_{\mathrm{FS}}, \label{eq:degeneracy} \\
&\mathrm{max}[p_{d1}^2(\boldsymbol{k})+p_g^2(\boldsymbol{k})]=1,
\end{align}
where $\braket{ A }_{\mathrm{FS}}$ denotes the averaged value of $A$ on the Fermi surface.
Within weak-coupling mean-field theory, the first condition in Eq.~(\ref{eq:degeneracy}) implies that the $d_{x^2-y^2}$-wave state and the $g_{xy(x^2-y^2)}$-wave state are degenerate.
As shown in Fig.~\ref{fig:figure2}(c), the $d_{x^2-y^2}+ig_{xy(x^2-y^2)}$-wave pair potential has the gap nodes along $k_a = \pm k_b$.
Lastly, we describe the $s+id_{xy}$-wave state by the pair potential
\begin{align}
&\Delta(\boldsymbol{k}) = \Delta_0 \{ p_{s}(\boldsymbol{k}) + i p_{d2}(\boldsymbol{k})\},\label{eq:pair4}\\
&p_{s}(\boldsymbol{k}) = c_{s},\\
&p_{d2}(\boldsymbol{k}) = c_{d2} \sin k_a a_{\parallel}\sin k_b a_{\parallel},
\end{align}
where $c_{s}$ and $c_{d2}$ are determined by the conditions
$\braket{p_{s}^2(\boldsymbol{k})}_{\mathrm{FS}} = \braket{p_{d2}^2(\boldsymbol{k})}_{\mathrm{FS}}$
and $\mathrm{max}[p_{s}^2(\boldsymbol{k})+p_{d2}^2(\boldsymbol{k})]=1$.
As shown in Fig.~\ref{fig:figure2}(d), the $(s+id_{xy})$-wave pair potential exhibits the largest amplitudes along $k_a = \pm k_b$, which appears to be inconsistent with the results of previously reported quasiparticle interference pattern experiments~\cite{madhavan_20}.
However, a microscopy theory predicting the $(s+id_{xy})$-wave states also considers a pair potential having the largest amplitudes along $k_a = \pm k_b$~\cite{kee_21}.
Thus, we adopt this theoretical prediction in the present paper.

In the following calculations, we fix the parameters as $k_Fa_{\parallel}=0.9 \pi$~\cite{singh_97} and $a_{\perp}/a_{\parallel}=3.3$~\cite{chmaissem_97},
which are reasonable to describe the Fermi surface of SRO.
In addition, we focus only on the low-transparency junctions with $z_0 = (m V_0/\hbar^2 k_F) = 2$.

\section{Differential conductance}%
%------------------------------------------------------------------------------
\begin{figure}[bbbb]
\begin{center}
\includegraphics[width=0.5\textwidth]{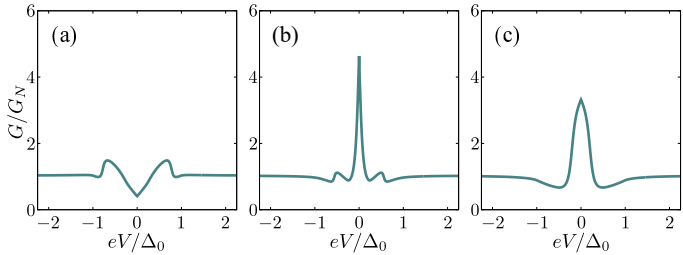}
\caption{Differential conductance of the $d_{zx}+id_{yz}$-wave superconductor as a function of the bias voltage.
In (a)--(c), we chose (a) $\theta=0$, (b) $\theta=0.125 \pi$, and (c) $\theta=0.25 \pi$.
}
\label{fig:figure4}
\end{center}
\end{figure}
%------------------------------------------------------------------------------
We first discuss the differential conductance of the $d_{zx}+id_{yz}$-wave state~\cite{kobayashi_15,tamura_17,suzuki_20} described by the pair potential in Eq.~(\ref{eq:pair1}).
Fig.~\ref{fig:figure3}(a) shows the differential conductance $G$ as a function of the bias voltage and the junction angle $\theta$.
The result is normalized by the normal conductance $G_N$ calculated by setting $\Delta_0=eV=0$.
At $\theta=0$, we obtain an M-shaped spectrum, as also shown in Fig.~\ref{fig:figure4}(a).
When the junction angle exceeds $\theta \approx 0.07 \pi$, the zero-bias conductance is enhanced.
Actually, the conductance spectrum at $\theta=0.125 \pi$ exhibits a sharp zero-bias peak, as also observed in Fig.~\ref{fig:figure4}(b).
We also observe two small humps at finite bias voltages, which are the remnants of the M-shaped spectrum at $\theta=0$.
The small humps at finite voltages disappear at $\theta \approx 0.25 \pi$.
Alternatively, we obtain a broad zero-bias peak structure, as also observed in Fig.~\ref{fig:figure4}(c), where the peak width at $\theta=0.25 \pi$ is much wider than that at $\theta=0.125 \pi$.
%------------------------------------------------------------------------------
\begin{figure}[bbbb]
\begin{center}
\includegraphics[width=0.5\textwidth]{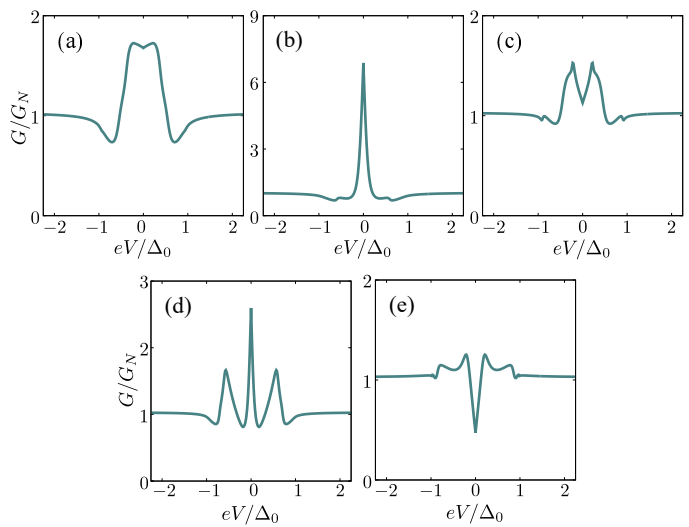}
\caption{Differential conductance of the $(d_{zx}+id_{yz})^{\prime}$-wave superconductor as a function of the bias voltage.
In (a)--(e), we chose (a) $\theta=0$, (b) $\theta=0.0625 \pi$, (c) $\theta=0.125 \pi$, (d) $\theta=0.1875 \pi$, and (e) $\theta=0.25 \pi$.}
\label{fig:figure5}
\end{center}
\end{figure}
%------------------------------------------------------------------------------

We next consider the $(d_{zx}+id_{yz})^{\prime}$-wave state described by the pair potential in Eq.~(\ref{eq:pair2}).
Fig.~\ref{fig:figure3}(b) shows $G$ as a function of the bias voltage and $\theta$.
At $\theta=0$, as shown in Fig.~\ref{fig:figure5}(a), the conductance spectrum exhibits a broad hump structure; a slight suppression around zero-bias voltage is also observed.
At $\theta \approx 0.0625\pi$, as shown in Fig.~\ref{fig:figure5}(b), the conductance spectrum shows a sharp zero-bias conductance peak
and nearly plateau regions on both sides of the peak.
At $\theta \approx 0.125\pi$, we observe an M-shaped hump structure (Fig.~\ref{fig:figure5}(c)).
At $\theta \approx 0.1875\pi$, the conductance spectrum shows three peaks (Fig.~\ref{fig:figure5}(d)).
At $\theta \approx 0.25\pi$, a sharp zero-bias dip appears in the conductance spectrum (Fig.~\ref{fig:figure5}(e)).

\begin{figure}[tttt]
\begin{center}
\includegraphics[width=0.5\textwidth]{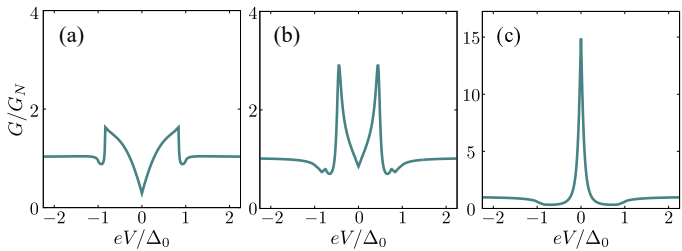}
\caption{Differential conductance of the $d_{x^2-y^2}+ig_{xy(x^2-y^2)}$-wave superconductor as a function of the bias voltage.
In (a)--(c), we chose (a) $\theta=0$, (b) $\theta=0.1875 \pi$, and (c) $\theta=0.25 \pi$.}
\label{fig:figure6}
\end{center}
\end{figure}

We here discuss the differential conductance of the $d_{x^2-y^2}+ig_{xy(x^2-y^2)}$-wave state described by the pair potential in Eq.~(\ref{eq:pair3}).
Fig.~\ref{fig:figure3}(c) shows $G$ as a function of the bias voltage and $\theta$.
As observed in Figs.~\ref{fig:figure6}(a) and \ref{fig:figure6}(b), the conductance spectra exhibit M-shaped structures except for $\theta \approx 0.25 \pi$.
In the spectra corresponding to $0.125\pi< \theta < 0.25\pi$, the width of the M-shaped hump decreases with increasing $\theta$.
We observe a sharp zero-bias conductance peak only at $\theta = 0.25 \pi$, as shown in Fig.~\ref{fig:figure6}(c).

%------------------------------------------------------------------------------
%------------------------------------------------------------------------------
%------------------------------------------------------------------------------
\begin{figure}[bbbb]
\begin{center}
\includegraphics[width=0.35\textwidth]{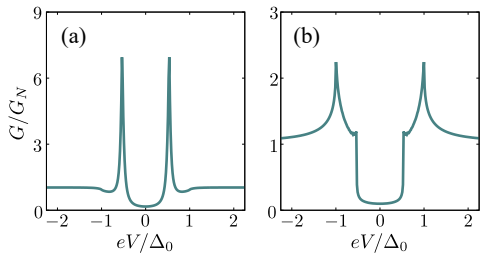}
\caption{Differential conductance of the $s+id_{xy}$-wave superconductor as a function of the bias voltage.
In (a) and (b), we chose (a) $\theta=0$ and (b) $\theta=0.25 \pi$.}
\label{fig:figure7}
\end{center}
\end{figure}
%------------------------------------------------------------------------------

Lastly, we discuss the differential conductance of the $s+id_{xy}$-wave state described by the pair potential in Eq.~(\ref{eq:pair4}).
We demonstrate $G$ as a function of the bias voltage and $\theta$ in Fig.~\ref{fig:figure3}(d).
As shown in Figs.~\ref{fig:figure7}(a) and \ref{fig:figure7}(b), the conductance spectrum exhibits gapped structures irrespective of $\theta$.
No enhancement of the conductance is observed at low bias voltages.

\section{Relevance to experiments}%
We here compare our theoretical results with previously reported experimental results.
The previous planar tunneling spectroscopy experiments~\cite{kashiwaya_11,yada_14} demonstrated
a broad hump peak structure, an M-shaped peak structure, and a two-step peak structure, where the sharp zero-bias peak formed on a small broad hump.
In these experiments, the fabricated devices had junction interfaces approximately perpendicular to the $a$-axis of SRO (i.e., $\theta \approx 0$).
As shown in Fig.~\ref{fig:figure5}, the $(d_{zx}+id_{yz})^{\prime}$-wave state leads to conductance spectra similar to those observed in the experiments.
Specifically, the broad hump structure is observed [Fig.~\ref{fig:figure5}(a)], along with the sharp peak structure with a plateau region [Fig.~\ref{fig:figure5}(b)] and the M-shaped hump structure [Fig.~\ref{fig:figure5}(c)] within $0 \leq \theta \leq 0.125 \pi$.
By contrast, other superconducting states show no enhancement in the low-bias conductance for $\theta \approx 0$.
Therefore, within our phenomenological single-band theory, the $(d_{zx}+id_{yz})^{\prime}$-wave state is the most plausible state for the superconductor SRO.
Nevertheless, our model for the $(d_{zx}+id_{yz})^{\prime}$-wave state is derived from rough assumptions regarding the long-range pairing terms.
In addition, the transport measurements along the $c$-axis~\cite{flouquet_09,firmo_15} result in gapped conductance spectra, suggesting the absence of the Andreev bound states at the top surface of SRO (i.e., the surface parallel to the $ab$-plane), which appears to be inconsistent with the $(d_{zx}+id_{yz})^{\prime}$-wave state (as well as the $d_{zx}+id_{yz}$-wave state)~\cite{kobayashi_15,tamura_17,suzuki_20}.
Therefore, a more conclusive result could be attained by examining the conductance spectra using additional microscopic models~\cite{agterberg_20(r)},
which is left as a future task.
From the experimental side, observing the tunnel conductance with various junction angles enables us to verify the realization of the $(d_{zx}+id_{yz})^{\prime}$-wave state more strictly.
Notably, a recent muon spin relaxation experiment under hydrostatic pressure~\cite{grinenko_21} also supports the realization of the $(d_{zx}+id_{yz})^{\prime}$-wave state (as well as the $d_{zx}+id_{yz}$-wave state), where the degeneracy of the two components of the order parameter is protected by symmetry.

We also note that there are several experimental findings that appear to contradict the aforementioned superconducting states.
For instance, recent specific heat measurements under uniaxial strain did not detect the split between the superconducting transition temperature and the temperature for the onset of broken time-reversal symmetry~\cite{mackenzie_21}; thus, this experimental result appears to contradict the two-component superconducting states.
Moreover, a recent Josephson current measurement suggests the realization of time-reversal invariant superconducting states~\cite{kashiwaya_19}.
Eventually, such inconsistencies might be resolved by future microscopic theories providing alternative interpretations of experimental results.
However, resolving these inconsistencies is beyond the scope of the present paper.
Alternatively, in the Supplemental Materials~\cite{supp_mat}, we discuss the tunneling spectroscopy of plausible superconducting states more exhaustively.
Specifically, we consider $d_{x^2-y^2}$-wave~\cite{simon_19},
nematic $d_{xz}$-wave~\cite{proust_20},
and spin-triplet helical $p$-wave states~\cite{kashiwaya_19,simon_19}.

\section{Conclusion}%
In summary, we discussed the tunneling spectroscopy of the three possible superconducting states of SRO, i.e., the $d_{zx}+id_{yz}$-wave, $d_{x^2-y^2}+ig_{xy(x^2-y^2)}$-wave, and $s+id_{xy}$-wave states.
Because microscopic models for SRO have not yet been established, we used the simple single-band model.
Nonetheless, our findings provide useful information for researchers in this field.
Specifically, we demonstrated that each superconducting state exhibits a characteristic pattern in the conductance spectrum with respect to the junction angle.
As long as we examine conductance spectra with a single junction angle, determining the pairing symmetry conclusively is difficult.
For instance, the broad hump spectrum observed in a planar tunneling experiment~\cite{kashiwaya_11} can be reproduced by various superconducting states other than the $(d_{zx}+id_{yz})^{\prime}$-wave state.
However, when a future study reports the conductance spectra of SRO with various junction angles,
the observed angular dependence will not be able to be reproduced by incorrect superconducting states.
That is, the characteristic pattern of the conductance spectrum with respect to the junction angle can function as a fingerprint for verifying the superconducting states of SRO.
We hope that this work motivates further tunneling spectroscopy experiments on SRO because such experiments have strong potential to shed light on the long-standing problem regarding its pairing symmetry.

\begin{acknowledgments}
We are grateful to M. Sigrist, S. Kashiwaya, and S. Tamura for fruitful discussions.
S.I. is supported by a Grant-in-Aid for JSPS Fellows (JSPS KAKENHI Grant No. JP21J00041).
S.-I. S. is supported by a Grant-in-Aid for JSPS Fellows (JSPS KAKENHI Grant No. JP19J02005).
Y. T. acknowledges support from a Grant-in-Aid for Scientific Research B (KAKENHI Grant No. JP18H01176 and No. JP20H01857)
and for Scientific Research A (KAKENHI Grant No. JP20H00131), and
Japan RFBR Bilateral Join Research Projects/Seminars number 19-52-50026.
This work is also supported by the JSPS Core-to-Core program Oxide Superspin International Network.
\end{acknowledgments}

\clearpage
\begin{center}
 \textbf{\large Supplemental Material for \\ ``Fingerprints of possible even-parity superconducting states in Sr$_2$RuO$_4$
\\ detected by planar tunneling spectroscopy''}\\ \vspace{0.3cm}
Satoshi Ikegaya$^{1,2}$, Shu-Ichiro Suzuki$^{2}$, Yukio Tanaka$^{2}$, and Dirk Manske$^{1}$\\ \vspace{0.1cm}
{\itshape $^{1}$Max-Planck-Institut f\"ur Festk\"orperforschung, Heisenbergstrasse 1, D-70569 Stuttgart, Germany\\
$^{2}$Department of Applied Physics, Nagoya University, Nagoya 464-8603, Japan}
\date{\today}
\end{center}

In this Appendix, we demonstrate the differential conductance of
the $d_{x^2-y^2}$-wave, nematic $d_{xz}$-wave, and spin-triplet helical $p$-wave states.

\subsection{$d_{x^2-y^2}$-wave state}
%------------------------------------------------------------------------------
\begin{figure}[hhhh]
\begin{center}
\includegraphics[width=0.45\textwidth]{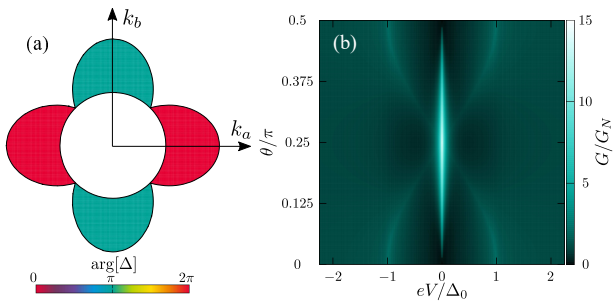}
\caption{(a) Pair potential of the $d_{x^2-y^2}$-wave state.
(b) Differential conductance as a function of the bias voltage and $\theta$.}
\label{fig:figure_ap1}
\end{center}
\end{figure}
%------------------------------------------------------------------------------
We describe the $d_{x^2-y^2}$-wave state by a pair potential,
\begin{align}
\Delta(\boldsymbol{k}) = \Delta_0 ( \cos k_b a_{\parallel} - \cos k_a a_{\parallel} ),
\end{align}
as also illustrated in Fig.~\ref{fig:figure_ap1}(a).
Fig.~\ref{fig:figure_ap1}(b) shows the differential conductance as a function of the bias voltage and the junction angle $\theta$. 
We observe the sharp zero-bias conductance peak except for $\theta=0$, where the largest value is taken at $\theta = 0.25 \pi$.
It is well known that the sharp zero-bias conductance peak of the $d_{x^2-y^2}$-wave superconductor is caused
by the mid-gap Andreev bound states~\cite{hu_94,tanaka_95}.

\subsection{Nematic $d_{xz}$-wave state}
%------------------------------------------------------------------------------
\begin{figure}[hhhh]
\begin{center}
\includegraphics[width=0.45\textwidth]{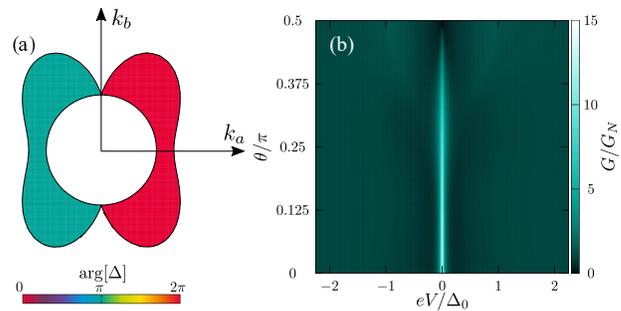}
\caption{(a) Pair potential of the nematic $d_{xz}$-wave state, where we choose $k_z = 0.5\pi$.
(b) Differential conductance as a function of the bias voltage and $\theta$.}
\label{fig:figure_ap2}
\end{center}
\end{figure}
%------------------------------------------------------------------------------
We describe the nematic $d_{xz}$-wave state by the pair potential
\begin{align}
\Delta(\boldsymbol{k}) = \Delta_0 \sin k_a a_{\parallel} \sin k_z a_{\perp} ,
\end{align}
as also shown in Fig.~\ref{fig:figure_ap2}(a).
As a result, as shown in Fig.~\ref{fig:figure_ap2}(b),
we observe the sharp zero-bias conductance peak except for $\theta = 0.5 \pi$, where the highest peak appears at $\theta=0$.

\subsection{helical $p$-wave state}
%------------------------------------------------------------------------------
\begin{figure}[hhhh]
\begin{center}
\includegraphics[width=0.45\textwidth]{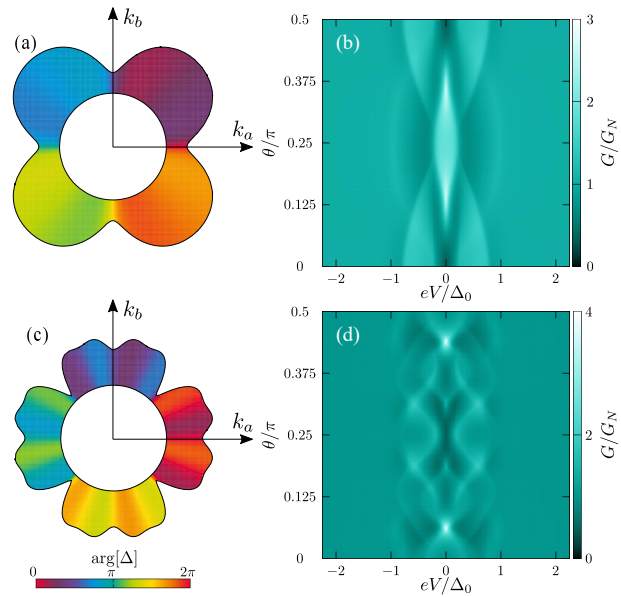}
\caption{(a) Pair potential of the helical $p$-wave state.
(b) Differential conductance of the helical $p$-wave superconductor as a function of the bias voltage and $\theta$.
(c) Pair potential of the helical $p^{\prime}$-wave state.
(d) Differential conductance of the helical $p^{\prime}$-wave superconductor as a function of the bias voltage and $\theta$.}
\label{fig:figure_ap3}
\end{center}
\end{figure}
%------------------------------------------------------------------------------
We describe a helical $p$-wave state by a pair potential,
\begin{align}
\Delta_{\pm}(\boldsymbol{k}) = \Delta_0 \left( \sin k_a a_{\parallel} \pm i \sin k_b a_{\parallel} \right),
\label{eq:pair5}
\end{align}
where $\Delta_{+}(\boldsymbol{k})$ and $\Delta_{-}(\boldsymbol{k})$ act on the electrons with different spins.
In the absence of spin-active perturbations such as Zeeman potentials and spin-orbit coupling potentials,
the differential conductance does not depend on the spin of electrons.
Namely, the superconducting state with $\Delta_{+}(\boldsymbol{k})$ and that with $\Delta_{-}(\boldsymbol{k})$ show the same conductance spectrum.
In Fig.~\ref{fig:figure_ap3}(a), we show the pair potential $\Delta_{+}(\boldsymbol{k})$.
We see that the pair potential in Eq.~(\ref{eq:pair5}) and the pair potential of the $d_{zx}+id_{yz}$-wave state in Eq.~(14) of the main text
have a same dependence with respect to the in-plane momenta (i.e., $k_a$ and $k_b$).
Consequently, the conductance spectra of the the helical $p$-wave state shown in Fig.~\ref{fig:figure_ap3}(b)
are very similar to that of $d_{zx}+id_{yz}$-wave state shown in Fig.~3(a) of the main text.

We also consider another pair potential for the helical $p$-wave state as,
\begin{align}
\Delta^{\prime}_{\pm}(\boldsymbol{k}) =\Delta_0 \{ p_a(\boldsymbol{k}) \pm i p_b(\boldsymbol{k}) \} , \label{eq:pair6}
\end{align}
with
\begin{align}
p_{a(b)}(\boldsymbol{k}) =& c_1 \sin k_{a(b)} a_{\parallel} \nonumber\\
&+ c_2 \cos  k_{b(a)} a_{\parallel} \times \sin k_{a(b)} a_{\parallel} \nonumber\\
&+ c_3 \sin 3k_{a(b)} a_{\parallel},
\end{align}
where we choose $c_1 = c_2 = c_3 = c$ and determine the coefficient $c$ by a condition of
$\mathrm{max}[p_{a}^2(\boldsymbol{k}) + p_{b}^2(\boldsymbol{k})]=1$.
In what follows, we refer to the superconducting state described by the pair potential in Eq.~(\ref{eq:pair6}) as the helical $p^{\prime}$-wave state.
The pair potential $\Delta^{\prime}_{+}(\boldsymbol{k})$ is shown in Fig.~\ref{fig:figure_ap3}(c).
We see that pair potential of the helical $p^{\prime}$-wave state and that of $(d_{zx}+id_{yz})^{\prime}$-wave state
have has a same dependence with respect to $k_a$ and $k_b$.
As a result, the conductance spectrum of the helical $p^{\prime}$-wave state shown in Fig.~\ref{fig:figure_ap3}(d)
and that of the $(d_{zx}+id_{yz})^{\prime}$-wave state in Fig.~3(b) of the main text become very similar.
Unfortunately, it seems to be difficult to distinguish
helical $p$-wave (helical $p^{\prime}$-wave) state and $d_{zx}+id_{yz}$-wave ($(d_{zx}+id_{yz})^{\prime}$-wave) solely by the planar tunneling spectroscopy.
Nevertheless, these two superconducting states has distinct differences in other aspects.
For instance, the helical $p$-wave state preserves time-reversal symmetry, while the $d_{zx}+id_{yz}$-wave state breaks the symmetry.
The $d_{zx}+id_{yz}$-wave state has horizontal line nodes, while the helical $p$-wave state does not have horizontal line nodes.
Therefore, these states would be clearly distinguished by other experiments.
\end{document}